\documentclass[twocolumn]{aastex631}
\usepackage{amsmath}
\usepackage{longtable}
\usepackage{afterpage}

\shorttitle{SN~2011aa}
\shortauthors{Dutta et al.}

\graphicspath{{./}{figures/}}

\begin{document}

\title{Can the violent merger of white dwarfs explain the slowest declining Type Ia supernova SN 2011aa?}

\correspondingauthor{Anirban Dutta}
\email{anirban.dutta@iiap.res.in, anirbaniamdutta@gmail.com}

\author[0000-0002-7708-3831]{Anirban Dutta}
\affiliation{Indian Institute of Astrophysics, II Block, Koramangala, Bangalore 560034, India}
\affiliation{Pondicherry University, R.V. Nagar, Kalapet, 605014, Puducherry, India}

\author[0000-0003-3533-7183]{G. C. Anupama}
\affiliation{Indian Institute of Astrophysics, II Block, Koramangala, Bangalore 560034, India}

\author{Nand Kumar Chakradhari}
\affiliation{School of Studies in Physics $\&$ Astrophysics, Pt. Ravishankar Shukla University, Raipur 492010, India}

\author[0000-0002-6688-0800]{D.K. Sahu}
\affiliation{Indian Institute of Astrophysics, II Block, Koramangala, Bangalore 560034, India}

\begin{abstract}

We present optical observations and Monte Carlo radiative transfer modeling of the Type Ia supernova SN~2011aa. With a $\Delta m_{15} (B)$ of $0.59 \pm 0.07$ mag and a peak magnitude $M_{\rm B}$ of $-19.30 \pm 0.27$ mag, SN~2011aa has the slowest decline rate among SNe Ia. The secondary maximum in $I$-band is absent or equally bright as the primary maximum.
The velocity of C\,{\sc ii} is lower than the velocity of Si\,{\sc ii}. This indicates either presence of C at lower velocities than Si or a line of sight effect. Application of Arnett's radiation diffusion model to the bolometric light curve indicates a massive ejecta $M_{\rm{ej}} ~ 1.8 - 2.6~M_{\odot}$. The slow decline rate and large ejecta mass, with a normal peak magnitude, are well explained by double degenerate, violent merger explosion model. The synthetic spectra and light curves generated with \texttt{SEDONA} considering a violent merger density profile match the observations. 

\end{abstract}


\section{Introduction} \label{sec:intro}

Type Ia supernovae (SNe Ia) result from thermonuclear explosions of white dwarf stars in binary systems (\citealt{1960ApJ...132..565H, 2019NatAs...3..706J}).
The rate of decline in normal type Ia SNe (0.85 $<$ $\Delta m_{15}(B)$ $<$ 1.70 mag) is correlated with the absolute magnitude in $B$-band \citep{1993ApJ...413L.105P}. The radioactive decay of $^{56}$Ni to $^{56}$Co and finally to $^{56}$Fe supplies the energy during the maximum of the light curve and its subsequent evolution (\citealt{1962PhDT........25P, 1969ApJ...157..623C, 2017suex.book.....B}). In addition, the efficiency with which $\gamma$-rays and positrons from the decay of $^{56}$Ni are trapped in the ejecta also plays an important role in the evolution of the light curve \citep{1997A&A...328..203C}. The luminosity also increases with more $^{56}$Ni produced in the explosion. This increased luminosity causes the ejecta to have a higher temperature. The opacity increases with temperature, and the diffusion timescales for the photons increases. This results in slower decline and broader light curves \citep{1996ApJ...472L..81H}. Hence, the decline rate versus absolute magnitude relation can also be interpreted as an opacity effect \citep{2012ApJ...753..105B}. While a majority of SNe Ia follow the luminosity decline rate relation, it is important to note that a good fraction of supernova events that are of thermonuclear origin do not follow this relation (\citealt{2017hsn..book..317T, 2016IJageePD..2530024M}). The over luminous super-Chandrasekhar SNe Ia (\citealt{2006Natur.443..308H, 2021ApJ...922..205A}) lie at the extreme end of the $\Delta m_{\rm 15}(B)$~-~$M_{\rm B}$ relation. They are slowly declining objects. The SNe Iax (SN~2002cx-like) are a peculiar class of thermonuclear explosions having low luminosity and low kinetic energy as compared to SNe Ia (\citealt{2003PASP..115..453L, 2022ApJ...925..217D}). 

In the proposed progenitor scenario for SNe Ia, the exploding white dwarf (WD) can have a non-degenerate star (single degenerate, SD) or another white dwarf (double degenerate, DD) as its binary companion. In the SD scenario, the white dwarf can accrete matter from a red-giant \citep{1992ApJ...397L..87M}, sub-giant/main-sequence \citep{1992A&A...262...97V} or a He star \citep{2010A&A...523A...3L}. In the DD case, violent merger of two similar mass WDs ($\sim$ 0.9 $M_{\odot}$) has been shown to give rise to subluminous type Ia SN explosion \citep{2010Natur.463...61P}. However, more massive primary white dwarfs, due to their higher densities will produce more $^{56}$Ni and Fe group elements and give rise to brighter SNe Ia \citep{2012ApJ...747L..10P}.

Using pre-explosion HST images \cite{2011Natur.480..348L} has ruled out He stars or luminous red giants as the companion of SN~2011fe. But main-sequence \citep{2011Natur.480..344N} or another WD as a companion cannot be ruled out. Observations of early UV emission in a thermonuclear supernova iPTF14atg \citep{2015Natur.521..328C} hinted towards collision of ejecta material with its companion, supporting a single degenerate scenario. The excess flux can also be explained by $^{56}$Ni in the outer layers \citep{2020A&A...642A.189M}. Some circumstellar mass can be formed from ejection of mass in tidal tails before the merger of two WDs. The interaction of the SN ejecta with the tidal tail ejecta produces signatures in X-ray/UV/optical \citep{2013ApJ...772....1R}. The persistent presence of hydrogen in the spectra of PTF11kx can be understood in terms of ejecta interacting with circumstellar mass indicating a non-degenerate companion (\citealt{2012Sci...337..942D, 2013ApJ...772..125S, 2017ApJ...843..102G}). The detection of [O\,{\sc i}] $\lambda\lambda$6300, 6364 in the nebular spectra of SN~2010lp indicates that oxygen is present close to the center which is predicted by violent merger scenario (\citealt{2013ApJ...775L..43T, 2013ApJ...778L..18K}). So, the very question of single/double degenerate progenitor still persists. The observed diversity in the explosions along with different models proposed to explain the diversity makes it important to study these systems.

In this Letter, we present optical observations and radiative transfer modeling with \texttt{SEDONA} of the spectra and light curves of a peculiar type Ia supernova SN 2011aa. SN~2011aa was discovered on 2011, February 6.3 in the galaxy UGC 3906 (PGC 021381) at $\alpha$ (J2000) = 07$^{h}$36$^{m}$42$^{s}$.63 and $\delta$ (J2000) = $+$74$^\circ$26$'$34$''$.80 \citep{2011CBET.2653....1P}. There is another nearby galaxy PGC 021386 with similar radial velocity as PGC 021381, making it a galaxy pair, with the SN located in between the two galaxies. \citealt{2011CBET.2653....3G} classified it as a type Ia supernova a few days before maximum light. The \textit{Swift}-UV observations of SN~2011aa were presented in \cite{2014ApJ...787...29B}. 
\cite{2015ApJS..220....9F} have catalogued the $JHK_{\rm s}$ photometry of SN~2011aa.

We organise the paper as follows - Section~\ref{Data Reduction} briefly discusses the observations and data reduction. The $UVOIR$-bolometric light curve and spectral evolution are analysed in Section~\ref{Light Curve}. We discuss possible explosion models in Section~\ref{Exp Models}. Section~\ref{Modeling} contains the Monte Carlo radiative transfer simulations and comparison of the synthetic spectra and light curves with the observed ones. Finally, we note the important features of SN~2011aa in Section~\ref{Discussions}.

\section{Data} \label{Data Reduction}

SN 2011aa was observed in imaging and spectroscopic mode with the 2m Himalayan Chandra Telescope\footnote{\url{https://www.iiap.res.in/?q=telescope_iao}} (HCT) of the Indian Astronomical Observatory (IAO), Hanle, India. Photometric monitoring  of supernova SN 2011aa began on 2011 February 08 (JD 245\,5601.33)  and continued until 2011 June 27 (JD 245\,5740.13). Spectroscopic observation of SN 2011aa was carried out during 2011 February 08 (JD 245\,5601.36) to 2011 April 29 (JD 245\,5681.18).

The images were obtained in Bessell $UBVRI$ filters and the spectra were obtained using grisms Gr~7 (3500--7800~\AA) and Gr~8 (5200--9100~\AA) with the  Himalayan Faint Object Spectrograph Camera (HFOSC), mounted on the HCT. Data are reduced in the same manner as described in \cite{2021MNRAS.503..896D}. Magnitudes are estimated using point spread function (PSF) fitting photometry and calibrated with respect to secondary standards in the field. 

The UltraViolet Imaging Telescope (UVOT) on board the {\it Swift} satellite observed SN 2011aa in three broad band optical filters $u$, $b$ and $v$ and three UV filters $uvw2$, $uvm2$ and $uvw1$, during  $\sim - 8$ d to $\sim$ +45 d with respect to $B$-band maximum. The {\it Swift}-UVOT data were downloaded from the \textit{Swift} archive and reduced using various modules in HEASoft (High Energy Astrophysics Software) following the methods of \cite{2008MNRAS.383..627P}, \cite{2009AJ....137.4517B} and \cite{2014MNRAS.443.1663C}.

\begin{figure}
\centering
\includegraphics[width=\columnwidth]{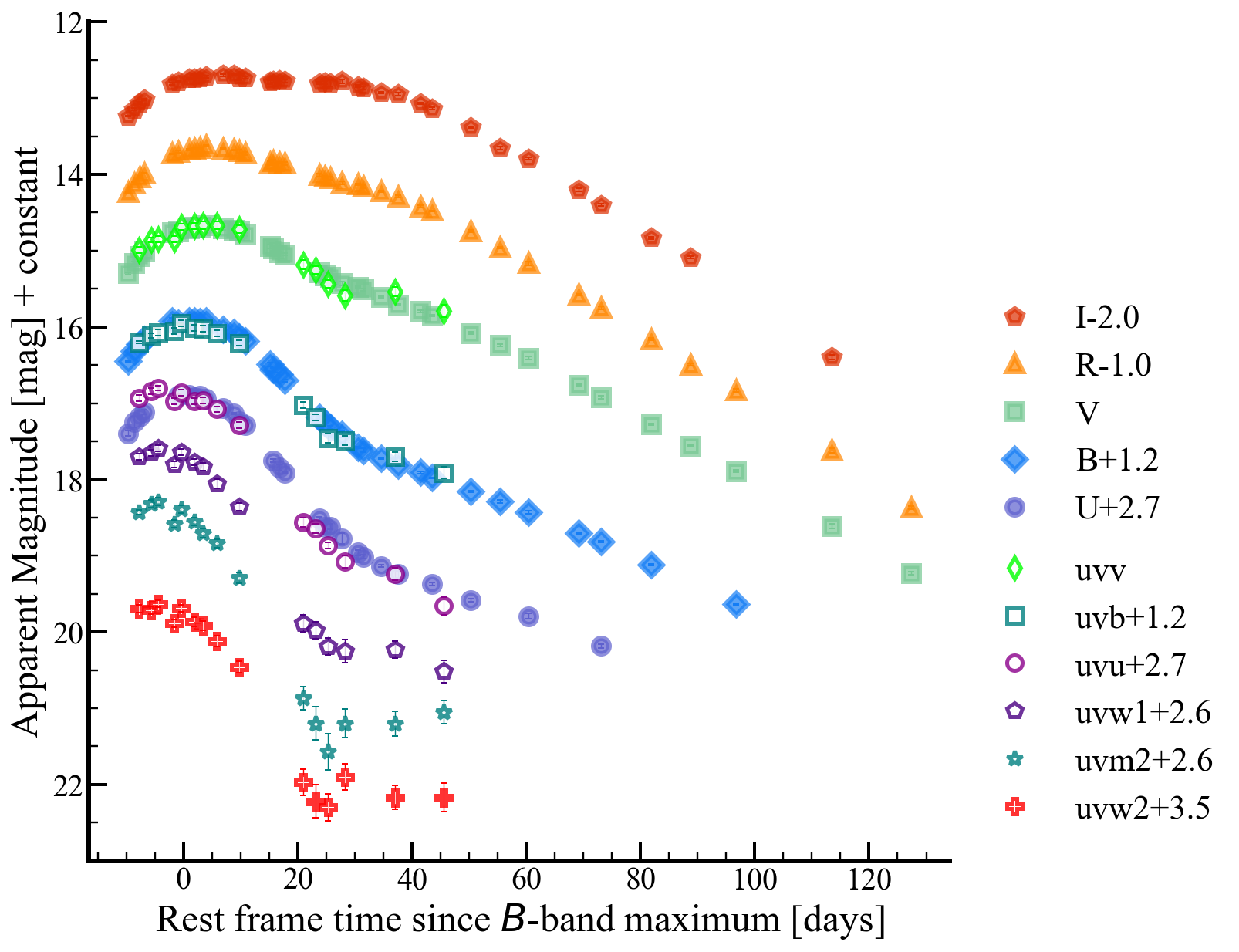}
\caption{{Optical $UBVRI$ and \it Swift}-UVOT light curves of SN~2011aa. The horizontal axis represent rest frame time since $B$-band maximum ($t$ - $t_{\rm max}$)/(1 + $z$). The magnitudes are given in \textit{Vega} system. The light curves have been vertically shifted by the amount indicated in the figure.} 
\label{Fig1}
\end{figure}

\section{Light Curve} \label{Light Curve}
\subsection{Light Curve Analysis}

The light curves of SN~2011aa in Bessell $UBVRI$ and $Swift$-UVOT $uvw2$, $uvm2$, $uvw1$, $u$, $b$ and $v$ bands are plotted in Fig.~\ref{Fig1}. The date of maximum, maximum magnitude are estimated by performing Gaussian process regression \citep{3569} with Matern kernel on the light curves, using the \texttt{gaussian\_process} package in \texttt{scikit-learn} \citep{scikit-learn}. Gaussian processes are supervised machine learning method designed to solve problems of regression. The errors are the standard deviation of 1000 such iterations. The peak in $B$-band occurred at JD 245\,5611.65$\pm$1.05 with a magnitude of $14.72 \pm 0.01$ mag. The $\Delta m_{15} (B)$ is estimated to be \textbf{$0.59 \pm 0.07$}, making it the slowest declining SN Ia. The peak in $U$-band occurred at JD~245\,5611.35$\pm$0.53 ($-$0.3~d) and that in $V$, $R$ and $I$-bands occurred at JD~245\,5615.52$\pm$0.44 ($+$3.8~d), 245\,5616.10$\pm$0.44 ($+$4.5~d) and 245\,5618.72$\pm$0.46 ($+$7.1~d). The $I$-band light curve does not show secondary peak which is a characteristic of a normal SN Ia. 

SN~2011aa exploded 20$''$.4 east and 10$''$.2 south of the center of UGC 3906 \citep{2011CBET.2653....1P}. The radial velocity of UGC 3906 corrected for Local Group infall onto Virgo is 3995 $\pm$ 20 km~$\rm s^{-1}$ \citep{2014A&A...570A..13M}. The calculated luminosity distance is 56.3 $\pm$ 0.3 Mpc and distance modulus 33.75 $\pm$ 0.27 mag, assuming $H_{\rm 0}$ = 70~km~$\rm s^{-1}$~$\rm Mpc^{-1}$, $\Omega_{\rm M}$ = 0.27 and $\Omega_{\rm \Lambda}$ = 0.73. The reddening is E$(B-V)$ = 0.0237 $\pm$ 0.0006 due to the dust in the Milky Way \citep{2011ApJ...737..103S}. However, in the near maximum spectrum of SN~2011aa we observe Na\,{\sc i}D ($\lambda$5890) absorption with a pseudo-equivalent width (pEW) of 0.41 $\pm$ 0.02 \AA\ due to interstellar medium in the Milky Way. Using the empirical relation E$(B-V)$ = 0.16 $\times$ pEW~(Na\,{\sc i}D) \citep{2003fthp.conf..200T}, we get E$(B-V)$ = 0.065 $\pm$ 0.003 mag. We do not detect any Na\,{\sc i}D at the redshift of the host galaxy, consistent with the location of SN~2011aa. The extinction in each band is estimated using \cite{1989ApJ...345..245C} with $R_{\rm V}$ = 3.1. The absolute magnitude in $B$-band is $-$19.30 $\pm$ 0.27 mag which is similar to normal SNe Ia. 

\subsection{Nickel mass and ejecta mass}

We obtain the $UVOIR$ bolometric light curve using the \textit{Swift}-UVOT, $UBVRI$ and $JHK_{\rm s}$ band magnitudes. The UV-contribution is 20$\%$ at $-$8.7~d and decreases to 3$\%$ at $+$27~d since $B$-band maximum. For the first two epochs before $-$8.7~d we do not have UV-magnitudes, so we add a 20$\%$ contribution to the optical. The NIR data coverage is from JD 245\,5616.17 to JD 245\,5657.19. The NIR contribution to the UV-optical luminosity is 10$\%$ at JD 245\,5616 ($+$4.4~d) and increases to 25$\%$ at JD 245\,5640 ($+$28.6~d). We assume a constant contribution of 10$\%$ from NIR before $+$4.4~d. The $UVOIR$ spectral energy distribution has been integrated from 1600 \AA\ to 24 800 \AA\ .

We fit the light curve with a modified radiation diffusion model \citep[See eqn. 9 of][]{2012ApJ...746..121C} upto 70 days since $B$-band maximum to obtain the parameters  $t_{exp}$ - the epoch of explosion, $M_{\rm{Ni}}$ - the $^{56}$Ni mass produced, $t_{\rm{lc}}$ - the light curve time scale and $t_\gamma$ - the gamma-ray leaking time scale. It is assumed that the initial radius of the progenitor is negligible compared to the expansion of the ejecta. A more realistic picture of the ejecta includes non-constant opacity, varying spatial distribution of the energy density ($^{56}$Ni mixing) (\citealt{2019ApJ...878...56K, 2019RNAAS...3..162K}).
We used the \texttt{emcee} package in \texttt{python} to find the posterior distribution and hence the upper and lower error limits. The details of the fitting procedure are described in \cite{2022ApJ...925..217D}.

The fit to the $UVOIR$ bolometric light curve gives $t_{exp}$ = $\rm 245\,5591.62^{+1.01}_{-1.41}$, $M_{\rm{Ni}}$ =  $0.87^{+0.06}_{-0.06}$ \(M_\odot\), $t_{\rm{lc}}$ = $16.82^{+1.68}_{-1.79}$ days and $t_\gamma$ = $55.42^{+3.00}_{-3.02}$ days.
The rise time in $B$-band from $t_{\rm exp}$ is $20.03^{+1.58}_{-1.48}$ days. Using a constant optical opacity $\kappa_{\rm opt}$ = 0.1 cm$^{2}$g$^{-1}$ and an expansion velocity $v_{\rm exp}$ = 12\,000 km s$^{-1}$ derived from the near-maximum spectrum, we get $M_{\rm{ej}}$ = $2.64^{+0.53}_{-0.56}$ \(M_\odot\) and a kinetic energy of explosion $E_{\rm{kinetic}}$ = $2.26^{+0.45}_{-0.48} \times 10^{51}$ erg. For a constant optical opacity of 0.15 cm$^{2}$g$^{-1}$, we get $M_{\rm{ej}}$ = $1.76^{+0.35}_{-0.37}$ \(M_\odot\) and a kinetic energy of explosion $E_{\rm{kinetic}}$ = $1.51^{+0.30}_{-0.32} \times 10^{51}$ erg. 
The $^{56}$Ni mass is within the range for normal SNe Ia, 0.09 - 0.87 $M_{\odot}$ with $\Delta m_{15} (B)$ between 0.8 - 1.9 mag \citep{2006A&A...460..793S}. The estimated $M_{\rm{ej}}$ is higher than expected for a normal Ia from a $M_{\rm ch}$ WD explosion. In order to understand the explosion mechanism and progenitor, we explore explosion models that can produce the ejecta and $^{56}$Ni mass as estimated by the analytical 1D model (See Sec.~\ref{Exp Models}). 

\subsection{Spectral evolution}

\begin{figure*}
\centering
\includegraphics[width=0.9\textwidth]{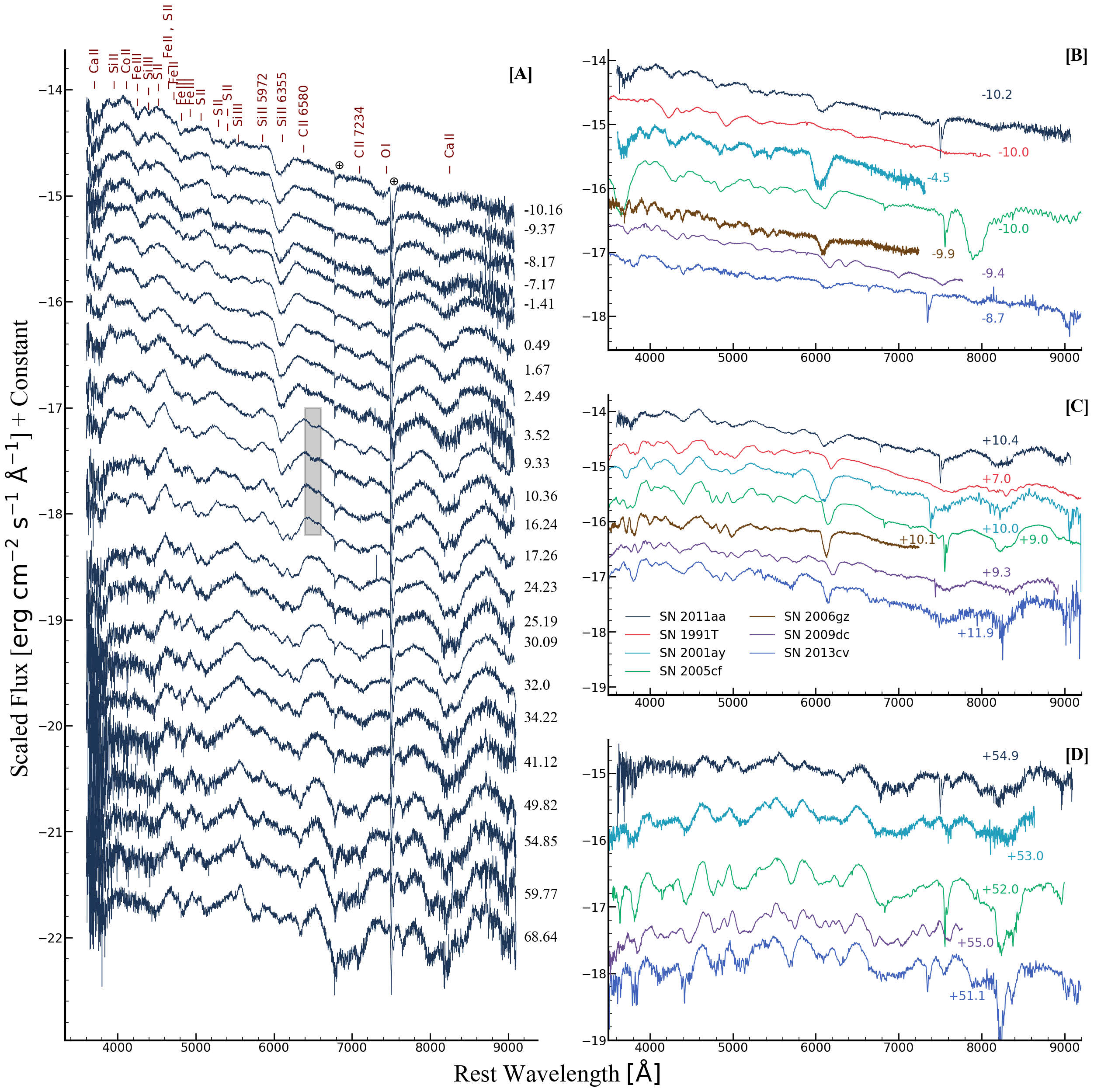}
\caption{Spectral evolution of SN~2011aa from $-$10.16 d to $+$68.6 d in the rest frame since $B$-band maximum. The important lines for the pre-mamimum spectra are marked in panel \textbf{[A]}. 
C\,{\sc ii} $\lambda$6580 region is shaded between 3.5 to 17.3~d since $B$-band maximum. The observed spectra are de-reddened and redshift corrected. Also shown is the comparison of SN~2011aa with luminous SN~1991T, peculiar SN~2001ay and SN~2013cv, normal SN~2005cf and super-Chandra SN~2006gz and SN~2009dc at pre-maximum \textbf{[B]}, near-maximum \textbf{[C]} and post-maximum \textbf{[D]} phase.}
\label{Fig2}
\end{figure*}

The spectral sequence of SN~2011aa is shown in Fig.~\ref{Fig2}\textbf{[A]}. The spectra at $-$10.2~d shows features due to Si\,{\sc ii}, Si\,{\sc iii}, Fe\,{\sc iii}, S\,{\sc ii}, C\,{\sc ii}, O\,{\sc i}, Ca\,{\sc ii}. SN~2011aa falls under the core-normal class of the Branch classification \citep{2006PASP..118..560B}. C\,{\sc ii} $\lambda$6580 and C\,{\sc ii}~$\lambda$7234 are seen in the spectra till $~$17~d since $B$-band maximum (Fig.~\ref{Fig2}[\textbf{A}]).

The ejecta velocity is measured by fitting a Gaussian function to the Si\,{\sc ii} $\lambda$6355 absorption feature. The velocity is $\sim$ 14\,000 km~$\rm s^{-1}$ at $-$10.2~d and decreases by 200 km~$\rm s^{-1}$ each day to reach 12\,200 km~$\rm s^{-1}$ around maximum. After maximum, the velocity evolves more slowly reaching about 12\,000 km~s$\rm ^{-1}$ at 3.5~day post maximum. The spectroscopic evolution is slow and forms a velocity plateau. This places SN~2011aa in the low velocity gradient group of the Benetti classification scheme \citep{2005ApJ...623.1011B}.  The velocity of C\,{\sc ii} $\lambda$6580 is 8980 $\pm$ 200 km~$\rm s^{-1}$ at $-$10.2~d and decreases to 3600 km~$\rm s^{-1}$ at $+$10.4~d. The velocity of  C\,{\sc ii}~$\lambda$7234 is 6650 $\pm$ 690~km~$\rm s^{-1}$ at $-$1.4~d and decreases to 3670 $\pm$ 180  km~$\rm s^{-1}$ at $+$10.4~d. The detonation wave proceeds faster at higher densities and the unburned material like C should be present at a higher velocity, lower density layer as compared to Si. The presence of C at lower velocities than Si may indicate that the photosphere at that epoch moves with the velocity of C, while Si layer is moving faster or ejecta asymmetries/clumping \citep{2011ApJ...732...30P}.

In Fig.~\ref{Fig2}[\textbf{B}], \textbf{[C]} and \textbf{[D]} we compare the spectra of SN~2011aa with SN~1991T \citep{1992AJ....104.1543F}, SN~2001ay \citep{2011AJ....142...74K}, SN~2005cf \citep{2009ApJ...697..380W}, SN~2006gz \citep{2007ApJ...669L..17H}, SN~2009dc \citep{2011MNRAS.412.2735T} and SN~2013cv \citep{2016ApJ...823..147C} at various epochs of evolution.
The spectroscopic evolution of SN~2011aa shows most differences with the comparison SNe in the early phase. It is quite different from bright SN~1991T-like objects that shows weak/no features due to intermediate mass elements in the pre-maximum spectra. Super-Chandra object like SN~2009dc ($M_{\rm B}$ = $-$20.22 mag with a $\Delta m_{15} (B)$ = 0.71~mag) produces 1.8 M$_{\odot}$  of $^{56}$Ni mass and 2.8 $M_{\odot}$ of ejecta. It shows strong absorption feature of C\,{\sc ii} (6580 \AA) in its spectra \citep{2011MNRAS.412.2735T} and is part of the shallow silicon (SS) group \citep{2006PASP..118..560B}. Another super-Chandra object SN~2006gz has velocity of $\sim$ 12 000 km~$\rm s^{-1}$ near maximum light and falls in the SS group. The velocity of C is more than Si in SN~2006gz. SN~2011aa does not show strong features due to Ca\,{\sc ii} as seen in normal SN~2005cf. SN~2013cv shows transitional nature between super-Chandra and normal SNe. It has lack of IGE's in the early phase spectra and persistent presence of C after maximum.

Now we discuss two objects which show some similarity with SN~2011aa in their light curve evolution. The peculiar SN~2001ay \citep{2011AJ....142...74K} exhibits slow decline ( $\Delta m_{15} (\rm B)$ = 0.68 mag) with $M_{\rm B}$ = $-$19.19 mag. The Si\,{\sc ii} $\lambda$6355 line is broad and its velocity evolution after maximum is quite rapid ($\sim$ 200 km~s$^{-1}$~d$^{-1}$) which makes it to fall under high velocity gradient group. This has been explained in terms of pulsational delayed detonation in a $M_{\rm ch}$ white dwarf \citep{2012ApJ...753..105B}. In the case of the peculiar ASASSN-15hy \citep{2021ApJ...920..107L} with a $\Delta m_{15} (\rm B)$ = 0.72 mag and $M_{\rm B}$ = $-$19.14 mag, the slow decline has been explained in terms of a degenerate core exploding inside a non-degenerate envelope. A large core mass is required in this case for understanding the broad light curves. The progenitor in this case is the merger of a white dwarf with the core of an asymptotic giant branch star. The observed spectral properties of ASASSN-15hy is more similar to SN~2009dc with low velocity near maximum (~8000 km~s$\rm ^{-1}$) and falls in the SS group. SN~2011aa is quite different in terms of spectroscopic evolution from both these objects.

\section{Explosion Models} \label{Exp Models}

In this section, we discuss various possible explosion models which may explain the observables of SN~2011aa. 

\subsection{Collision of white dwarfs}

Collision between two white dwarfs can occur in dense stellar environments like the core of globular clusters. \cite{2009MNRAS.399L.156R} calculated about 10-100 white dwarf collisions per year at redshift $\le$ 1. They calculated the explosion parameters for the collision of two equal mass white dwarfs (0.6 $M_{\odot}$) for different impact parameters. The $^{56}$Ni mass is maximum ($\sim$ 0.4 $M_{\odot}$) for a head on collision and decreases with increasing impact parameter. 
This $^{56}$Ni mass is not consistent with SN~2011aa. Collision of higher mass white dwarfs ($\sim$ 0.9 $M_{\odot}$) with low impact parameter can produce brighter events that broadly follow the Phillips relation \cite[See Fig.~3 of][]{2009ApJ...705L.128R} with some dependence on viewing angle. Using similar and dissimilar masses of the colliding white dwarfs \cite{2013ApJ...778L..37K} showed that SN Ia explosions are produced with $^{56}$Ni mass in the range of 0.1 $-$ 1.0 $M_{\odot}$. Both the work of \cite{2009ApJ...705L.128R} and \cite{2013ApJ...778L..37K} showed that the ejecta structure is stratified with C and O in the outer layers, intermediate mass elements (IME's) in the inner layers and Fe group elements (IGE's) in the innermost layers caused as a result of detonation.

\subsection{Rotating white dwarfs}

The maximum mass of a non-rotating, inert WD is $\sim$ 1.4 $M_{\odot}$ (M$_{\rm ch}$). Carbon-oxygen white dwarfs having differential rotation can support mass exceeding the Chandrasekhar limit \citep{2005A&A...435..967Y}. If the rotating white dwarf is having a non-degenerate companion, the maximum possible mass that the accreting white dwarf can reach by accretion is 2 $M_{\odot}$ \citep{2000A&A...362.1046L}. The efficiency of the angular momentum gain and the loss will decide the final SN Ia explosion. We discuss explosion models based on rapidly rotating C-O white dwarfs under hydrostatic equilibrium. 

\cite{2018A&A...618A.124F} constructed white dwarfs within mass range of 1.6 - 2.0 $M_{\odot}$ and angular momentum between 0.9 - 2.2 $\times$ 10$^{50}$ g~cm$^{2}$~$\rm s^{-1}$. Different explosion models like prompt detonation, delayed detonation (DDT) and pure deflagration models were tested. The prompt detonation model produces 1.44 $M_{\odot}$ of $^{56}$Ni. In this model, the intermediate mass elements (IME) produced is extremely low. In the delayed detonation models, an initial deflagration develops which transitions to a detonation due to the Zel'dovich gradient mechanism (mixing of hot ash with cold fuel under gravity). The mass of $^{56}$Ni varies between 1.06 - 1.45 $M_{\odot}$ based on the white dwarf mass. All the DDT models show asymmetric ejecta structure. The models are luminous with peak $B$-band absolute magnitude around $\sim$ $-$ 20 mag. Of particular interest is the model AWD1, in which 1.06 M$_{\odot}$ of $^{56}$Ni is produced. This value is close to the $^{56}$Ni mass estimate of SN~2011aa. But the model spectra is bluer with high blueshift of Si\,{\sc ii} line. The model produces huge amount of IGE's (1.31 M$_{\odot}$) inconsistent with the observed spectra of SN~2011aa. The ejecta structure is also highly stratified with very less unburned C/O in the outer layers due to detonation. We show the angle averaged spectra of AWD1 model \citep{2018A&A...618A.124F} compared with SN~2011aa in Fig.~\ref{Fig3}.

\subsection{Violent Merger}

Three-dimensional simulations of violent merger of two C-O white dwarfs of dissimilar masses (0.9 $M_{\odot}$ and 1.1 $M_{\odot}$) can explain the normal type Ia explosions \citep{2012ApJ...747L..10P}. The material from the secondary is violently accreted onto the primary. The material is then compressed and heated up on the surface of the primary where carbon burning is ignited. At a density ($\rho$) 2 $\times$ $10^{6}$ g $\rm cm^{-3}$ and at a temperature (T) greater than 2.5 $\times$ $10^{9}$ K a detonation occurs. The detonation flame propagates through the final merged object and burns the material. The energy released unbinds the object. The total ejected mass of this model ($\sim$ 1.95 M$_{\odot}$) is comparable to that of SN~2011aa estimated using Arnett's model. This model leads to an asymmetric explosion, hence the observables have a line of sight dependence. In $B$-band the peak magnitude varies between $-$19.5 to $-$18.7 mag and the angle averaged magnitude is $-$19.0 mag. Similarly, the $\Delta m_{15}$($B$) varies between 0.5 and 1.4 mag with the mean value being 0.95 mag. The estimated values of the observables of SN~2011aa lie well within the range predicted by the violent merger model. The primary white dwarf is burned and its ashes expand while the unburned and incompletely burned material (C, O, Ne, Mg) from the secondary white dwarf resides near the center of the ejecta. Hence, the presence of C/O at lower velocities may be crucial in determining the explosion scenario. We show the angle averaged spectra of violent merger model from \cite{2012ApJ...747L..10P} in Fig.~\ref{Fig3}. The spectra shows some similarity in velocities and line strengths to SN~2011aa. This encourages us to use the violent merger model varying the abundances at different velocities to accurately model the line strengths and velocities and get estimates of mass of different elements synthesized. We used \texttt{SEDONA} with violent merger density profile to model the observed spectra and light curve.

\section{Modeling of the spectra and light curves} \label{Modeling}

We use the multi-dimensional Monte Carlo radiative transfer code \texttt{SEDONA} \footnote{\url{https://github.com/dnkasen/pubsed}} \citep{2006ApJ...651..366K} to simulate the spectral and light curve evolution of SN~2011aa. The assumptions of the code are homologous expansion, Sobolev approximation and local thermodynamic equilibrium (LTE). \texttt{SEDONA} takes into account the energy deposition from the radioactive decay of $^{56}$Ni - $^{56}$Co - $^{56}$Fe. All the lines are treated in the expansion opacity formalism using the two level atom approach. The parameter $\rm \epsilon$ controls the probability of redistribution of radiation. 

\begin{figure*}
\centering
\includegraphics[width=\textwidth]{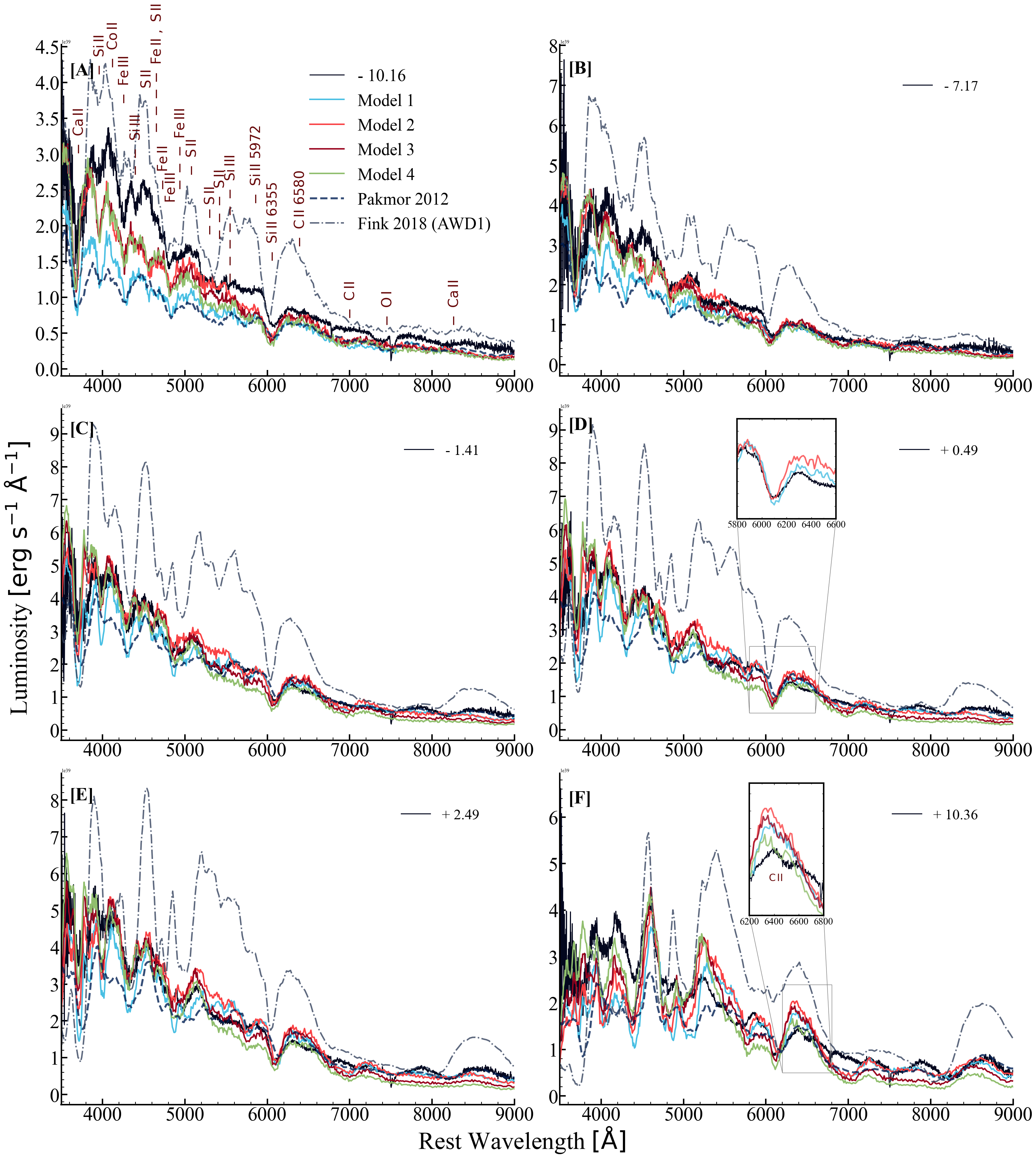}
\caption{Spectral evolution of SN~2011aa from $-$10.16 d to $+$10.36 d since rest frame time in $B$-band maximum. We show the model spectra calculated using \texttt{SEDONA}. Also shown are the angle averaged spectra of accreting white dwarf model (AWD1) from \cite{2018A&A...618A.124F} and violent merger model from \cite{2012ApJ...747L..10P}. These are taken from the \texttt{HESMA} database. The important lines in the early phase are marked in \textbf{[A]}. The observed spectra are de-reddened and redshift corrected.}
\label{Fig3}
\end{figure*}

For all the models in this work, we use the spherical one-dimensional ejecta structure and the one dimensional angle averaged density profile of the violent merger model \citep[merger\_2012\_11+09,][]{2012ApJ...747L..10P}. We consider 5 $\times$ 10$^{5}$ particles, 1000 frequency bins between 10$^{14}$ Hz and 10$^{16}$ Hz, 100 logarithmically spaced time steps and start the simulation at 2 days since explosion and evolve the models till 60 days. It is to be noted that for the purpose of resolving the most prominent features in the spectra, 5 $\times$ 10$^{5}$ particles is sufficient although increasing the number of particles will increase the resolution at the expense of computation time. We do not consider any best fitting technique in this work.

Fig.~\ref{Fig3} shows the spectral evolution of SN~2011aa along with synthetic spectrum generated with \texttt{SEDONA}. We consider C, O, Ne, Na, Mg, Si, S, Ca, $^{54}$Fe, $^{56}$Ni and $^{58}$Ni in the models. We consider four models in this work based on the violent merger density profile. The integrated ejected mass is 1.95 $M_{\odot}$ with a kinetic energy of 1.7 $\times$ 10$^{51}$ erg for the models.

In the first model (Model 1), we use the abundance of the violent merger model \citep{2012ApJ...747L..10P} which produces 0.62 $M_{\odot}$ of $^{56}$Ni. In the model, C and O are present in the very inner layers ($<$ 1000 km~s$^{-1}$). This is due to the fact that the burning is not complete for the lesser dense secondary white dwarf and unburned elements dominate near the center after the ashes of the primary has expanded. $^{56}$Ni is present upto 11\,600 km~s$^{-1}$. The layers between 10\,000 - 15\,000 km~s$^{-1}$ are dominated by Si, S, Mg. The outer layers above 20\,000 km~s$^{-1}$ are mostly C and O. 

The $^{56}$Ni mass is lower than the mass estimated for SN~2011aa using Arnett model fit to the $UVOIR$ light curve. In this model, the mass of unburned elements (C, O and Ne) is 0.82 $M_{\odot}$, intermediate mass elements (Mg, Si, S, Ca) is 0.47 $M_{\odot}$ and Fe-group elements ($^{54}$Fe, $^{56}$Ni and $^{58}$Ni) is 0.67 $M_{\odot}$. 
The model spectrum at $-$10.16~d since $B$-band maximum is quite red and does not reproduce the observed continuum (Fig.~\ref{Fig3}\textbf{[A]}). The Fe\,{\sc iii}, S\,{\sc ii} and Mg\,{\sc ii} lines are very strong in the model (Fig.\ref{Fig3}\textbf{[C]} and \textbf{[D]}) indicating an over abundance of these elements. Near the maximum ($+$0.49~d) Si\,{\sc ii} $\lambda$6355 and C\,{\sc ii} $\lambda$6580 lines are reproduced well. In the spectrum taken $+$10.4~d after maximum, we see C\,{\sc ii} $\lambda$6580 absorption feature (see inset of Fig.~\ref{Fig3}\textbf{[F]}). In normal SNe Ia, C is seen in the spectra during pre-maximum to maximum phases. However, in super-Chandrasekhar objects like SN~2009dc \citep{2011MNRAS.412.2735T}, C can be seen in the post-maximum phases also. The detection of C in this phase indicates that C is present in the inner layers. This supports our argument of a violent merger scenario. At $+$10.4~d, the region below 5000 \AA\ gets redder. This is mostly because of line blocking by Fe group elements.

\begin{figure}
\centering
\includegraphics[width=\columnwidth]{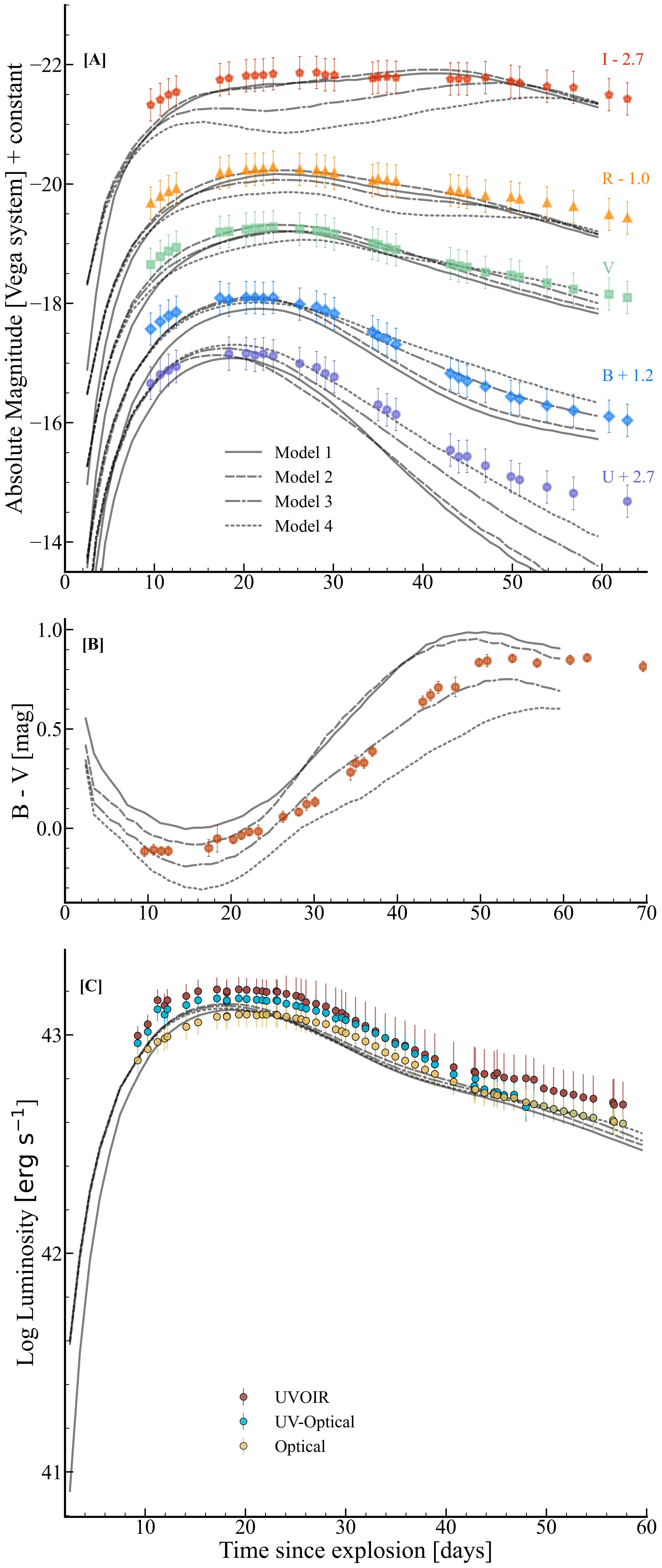}
\caption{\textbf{[A]} The $U$, $B$, $V$, $R$ and $I$ band light curves of SN~2011aa along with the synthetic light curves generated using \texttt{SEDONA}. From the model spectral energy distribution we calculate the light curves in \textit{AB} system and convert to \textit{Vega} system using values given in \cite{2007AJ....133..734B}. The horizontal axis is time since explosion (JD 245\,5591.62). \textbf{[B]} The $B-V$ color evolution of SN~2011aa plotted along with the \texttt{SEDONA} models. \textbf{[C]} The optical, UV-optical and $UVOIR$ bolometric light curves is shown along with the models.}
\label{Fig4}
\end{figure}

To account for the $^{56}$Ni mass estimate from the analytical light curve model (0.87 $\pm$ 0.06 $M_{\odot}$), we increased the mass fraction of $^{56}$Ni at the expense of Mg and S between 10\,000 - 16\,000 km~$\rm s^{-1}$ in the second model (Model 2). Between 8000 - 10\,000 km~$\rm s^{-1}$, we decreased S and increased $^{56}$Ni. We also reduced the mass fraction of $^{54}$Fe between 8000 - 16\,000 km~$\rm s^{-1}$. This allows to reproduce the line strengths of Fe\,{\sc iii} and S\,{\sc ii} in the models (Fig.~\ref{Fig3}\textbf{[C]} and \textbf{[D]}). Near the maximum, the red wing of the Si\,{\sc ii} $\lambda$6355 feature is stronger than the first model (see inset of Fig.~\ref{Fig3}\textbf{[D]}). This is because of the higher ionization at lower velocities caused by an increased $^{56}$Ni abundance. In this model, the $^{56}$Ni mass is 0.68 $M_{\odot}$. Further increasing the $^{56}$Ni mass makes the red wing of Si\,{\sc ii} $\lambda$6355 even stronger. Hence we do not consider increasing $^{56}$Ni further. 

\begin{table*}
\centering
\caption{\texttt{SEDONA} model features}
\label{tab:Models}
\begin{tabular}{lcccccc}
\hline\hline
Model Name & Density$^{*}$   &    Abundance    &  $\epsilon$  & $\Delta m_{15} (B)$ (mag) & $M_{\rm B}$ (mag)  & $^{56}$Ni ($M_{\odot}$)\\
\hline\hline  
Model 1  & violent merger &  violent merger & 1.0    &  0.96  & $-$19.20 & 0.62 \\
Model 2  & violent merger &  modified abundance &  1.0    &  0.96  & $-$19.39 & 0.68 \\
         &                &  $+$ increased $^{56}$Ni &    &  &    &  \\
Model 3  & violent merger &  same as Model 2 & 0.5    & 0.77  & $-$19.35  & 0.68 \\
Model 4  & violent merger & same as Model 2 & 0.3   & 0.59   & $-$19.31  & 0.68 \\
\hline
\multicolumn{2}{l}{$^{*}$\footnotesize{One dimensional density of the violent merger model}.}
\end{tabular}
\end{table*}

The region below 5000 \AA\ is redder than observed in this model also. For both the models, the redistribution probability $\epsilon$ is 1.0 which means absorption dominated treatment of the redistribution of radiation. Following a source function the absorption by a line is followed by a re-emission in another frequency. This acts as fluorescence. We construct two more models varying the $\epsilon$ parameter. In these two models the abundances are same as the second model. In the third model (Model 3), with $\epsilon$=0.5, meaning equal probability for scattering and absorption, we find slight improvement in reproducing the blueward flux at $+$10.4~d over the second model. In the fourth model (Model 4) $\epsilon$=0.3, there is severe flux depression in the region around 5200 - 6000 \AA\ and an increase in flux in the bluer region. This means in this model, the blue photons are scattered out of the ejecta more rather than being absorbed.

The light curves in each bandpass is obtained from the synthetic spectral energy distribution by convolving it with the HFOSC filter response. In Fig.~\ref{Fig4}\textbf{[A]}, we plot the observed light curves of SN~2011aa in $U$, $B$, $V$, $R$ and $I$ along with the model light curves from \texttt{SEDONA} simulation. Model 1 (solid curve in Fig.~\ref{Fig4}\textbf{[A]}), under predicts the magnitude in all bands. This is due to lower $^{56}$Ni mass in the model ejecta. In Model 2 (dashed), we increase the $^{56}$Ni mass and find that the peak in $B$-band is close ($-$19.39 mag) to the observed value ($-$19.30 $\pm$ 0.27 mag). But the decline rate is 0.96 mag which is faster than the observed light curve (0.59~mag). The model with $\epsilon$=0.5 (dash-dot), the peak magnitude in $B$-band is $-$19.35~mag with a decline rate of 0.77 mag. In the fourth model $\epsilon$=0.3 (dotted), the peak magnitude is $-$19.31 mag with a decline rate of 0.59 mag in $B$-band. But the flux in $I$-band is severely under predicted. The absence or equally bright secondary maximum is reproduced by our violent merger models (particularly Model 1 $\&$ Model 2).
The extinction corrected $B-V$ color is plotted and compared with the models in Fig.~\ref{Fig4}\textbf{[B]}. Model 1 is redder and Model 4 is bluer than the observed values throughout. Increasing $^{56}$Ni makes Model 2 bluer than Model 1 around 10-20 days since explosion, however this model becomes redder after 30 days. Model 3 predicts the color evolution better around 30-50 days. The optical, UV-optical and $UVOIR$ bolometric luminosity are plotted in Fig.~\ref{Fig4}\textbf{[C]} along with the \texttt{SEDONA} models. Due to less $^{56}$Ni in Model 1, the luminosity in the early phase is dimmer than the other models. Increasing $^{56}$Ni and distributing to outer layers ($\sim$ 16 000 km~$\rm s^{-1}$) increases the flux in the early times. The effect of $\epsilon$ is not seen in the model bolometric light curves.
The model parameters are listed in Table~\ref{tab:Models}.

Considering the violent merger density profile, we find that all the models can reproduce the observed light curves and spectra fairly reasonably. Using non-LTE calculations in one-dimension, \cite{2021ApJ...909L..18S} showed that there is a reduction in Fe\,{\sc ii} line blanketing and Ca\,{\sc ii} emission after peak which increases the magnitude in $U$, $B$-bands and decreases in $I$-band while the $V$-band is mostly unaffected. A detailed three dimensional, non-LTE consideration of the radiation diffusion treating fluorescence for each line separately may give better match to the observables of SN~2011aa.  

\section{Discussions} \label{Discussions}

For normal SNe Ia, the average decline rate $\Delta m_{15} (B)$ is 1.2 mag with a peak $B$-band magnitude of $-$19.3 $\pm$ 0.1 mag \citep{2005ApJ...623.1011B}. They show presence of a secondary maximum in $I$-band light curve. The average expansion velocity near maximum is 10\,600 $\pm$ 400 km~s$\rm ^{-1}$ \citep{2009ApJ...699L.139W}. The favoured explosion mechanisms for the normal SNe Ia, is the delayed detonation \citep{2007Sci...315..825M} in a $M_{\rm ch}$ or detonation in a sub-$M_{\rm ch}$ WD \citep{2010ApJ...714L..52S}.

SN~2011aa is the slowest declining type Ia SN with a $\Delta m_{15} (\rm B)$ is 0.59 $\pm$ 0.07 mag and peak magnitude of $-$19.30 $\pm$ 0.27 mag. Analytical models indicate $^{56}$Ni mass is 0.87~$M_{\odot}$ and ejected mass is between 1.76 - 2.64 $M_{\odot}$. The secondary maximum in $I$-band is as bright as the primary. The Si\,{\sc ii} velocity evolution shows a plateau after maximum. The velocity plateau can be explained in terms of a merger scenario where there is a C-O white dwarf inside an extended envelope \citep{1993A&A...270..223K}. This extended envelope is formed by the destruction of the secondary white dwarf. A detonation shock wave propagating outwards will collide with this low density envelope and an inward shock wave will cause deceleration of the outward moving material. The duration of the plateau observed in the velocity will be dependent on the mass of the envelope. Due to the interaction of the shock wave with the envelope, there will be density and pressure gradients which will cause mixing of the materials in the ejecta. The velocity of C\,{\sc ii} $\lambda$6580 and C\,{\sc ii} $\lambda$7234 is less than the velocity of Si\,{\sc ii} $\lambda$6355. This can be due to C\,{\sc ii} present in the inner layers or clumping/ line of sight effect. SN~2011aa shows slow decline similar to that seen in super-Chandra objects, however, the spectral evolution is not similar to either normal or super-Chandra objects. The merger scenario produces similar $^{56}$Ni mass as compared to normal SNe Ia and more ejecta mass ($\sim$ 1.95 $M_{\odot}$). So, the peak luminosity is similar to normal Ia and with more ejecta the diffusion time for the photons is large making the decline rate slower.
In this work, we have demonstrated that SN~2011aa with slower decline rate but normal peak magnitude can be explained by violent merger of white dwarfs.

\section{Acknowledgments}

We thank the anonymous referee for very useful comments and suggestions. We thank the staff of IAO, Hanle and CREST, Hosakote that made the observations possible. The facilities at IAO and CREST are operated by the Indian Institute of Astrophysics, Bangalore. We also thank the observers of HCT who shared their valuable time for Target of Opportunity (ToO) observations during the initial follow up. This work has made use of the NASA Astrophysics Data System\footnote{\url{https://ui.adsabs.harvard.edu/}} ({\sc ADS}), the NASA/IPAC extragalactic database\footnote{\url{https://ned.ipac.caltech.edu/}} ({\sc NED}) and NASA/IPAC Infrared Science Archive ({\sc IRSA})\footnote{\url{https://irsa.ipac.caltech.edu/applications/DUST/}} which is operated by the Jet Propulsion Laboratory, California Institute of Technology. We acknowledge, Weizmann Interactive Supernova Data REPository\footnote{\url{https://wiserep.weizmann.ac.il/}} (WISeREP), \citep{2012PASP..124..668Y}. This research has made use of the data obtained from the High Energy Astrophysics Science Archive Research Center (HEASARC) \footnote{\url{https://heasarc.gsfc.nasa.gov}} a facility of the Astrophysics Science Division at NASA/GSFC and of the Smithsonian Astrophysical Observatory’s High Energy Astrophysics Division.
This work made use of the Heidelberg Supernova Model Archive ({\sc hesma})\footnote{\url{https://hesma.h-its.org}}. This research has made use of the Spanish Virtual Observatory\footnote{\url{https://svo.cab.inta-csic.es}} \citep{2020sea..confE.182R}. We acknowledge the usage of the HyperLeda database \footnote{\url{http://leda.univ-lyon1.fr}} \citep{2014A&A...570A..13M}. We acknowledge the Weizmann Interactive Supernova Data Repository\footnote{\url{https://wiserep.weizmann.ac.il/}} (WISeREP, \cite{2012PASP..124..668Y}

The analysis has made use of the following software and packages - (i) {\small \it{Image Reduction and Analysis Facility}} (\texttt{IRAF}), \cite{1993ASPC...52..173T}. (ii) \texttt{PyRAF}, \cite{2012ascl.soft07011S}. (iii) \texttt{NumPy}, \cite{van2011numpy}, (iv) \texttt{Matplotlib}, \cite{Hunter:2007}, (v) \texttt{Scipy}, \cite{2020SciPy-NMeth}, (vi) \texttt{pandas}, \cite{reback2020pandas}, (vii) \texttt{Astropy}, \cite{2013A&A...558A..33A}, (viii) \texttt{emcee}, \cite{2013ascl.soft03002F}, (ix) \texttt{scikit-learn}, \cite{scikit-learn} (x) \texttt{SEDONA}, \cite{2006ApJ...651..366K}. 

\section{Data and Model availability}

Reduced spectra presented in this paper will be made available in wiserep archive. Models and reduced spectra are available in github (\url{https://github.com/Knights-Templars/SN2011aa}). Photometric data is given as data behind Fig.~\ref{Fig1}.

\bibliographystyle{aasjournal}
\bibliography{2011aa_SNIa}

\begin{thebibliography}{}
\expandafter\ifx\csname natexlab\endcsname\relax\def\natexlab#1{#1}\fi
\providecommand{\url}[1]{\href{#1}{#1}}
\providecommand{\dodoi}[1]{doi:~\href{http://doi.org/#1}{\nolinkurl{#1}}}
\providecommand{\doeprint}[1]{\href{http://ascl.net/#1}{\nolinkurl{http://ascl.net/#1}}}
\providecommand{\doarXiv}[1]{\href{https://arxiv.org/abs/#1}{\nolinkurl{https://arxiv.org/abs/#1}}}

\bibitem[{{Ashall} {et~al.}(2021){Ashall}, {Lu}, {Hsiao}, {Hoeflich},
  {Phillips}, {Galbany}, {Burns}, {Contreras}, {Krisciunas}, {Morrell},
  {Stritzinger}, {Suntzeff}, {Taddia}, {Anais}, {Baron}, {Brown}, {Busta},
  {Campillay}, {Castell{\'o}n}, {Corco}, {Davis}, {Folatelli}, {F{\"o}rster},
  {Freedman}, {Gonzal{\'e}z}, {Hamuy}, {Holmbo}, {Kirshner}, {Kumar}, {Marion},
  {Mazzali}, {Morokuma}, {Nugent}, {Persson}, {Piro}, {Roth}, {Salgado},
  {Sand}, {Seron}, {Shahbandeh}, \& {Shappee}}]{2021ApJ...922..205A}
{Ashall}, C., {Lu}, J., {Hsiao}, E.~Y., {et~al.} 2021, \apj, 922, 205,
  \dodoi{10.3847/1538-4357/ac19ac}

\bibitem[{{Astropy Collaboration} {et~al.}(2013){Astropy Collaboration},
  {Robitaille}, {Tollerud}, {Greenfield}, {Droettboom}, {Bray}, {Aldcroft},
  {Davis}, {Ginsburg}, {Price-Whelan}, {Kerzendorf}, {Conley}, {Crighton},
  {Barbary}, {Muna}, {Ferguson}, {Grollier}, {Parikh}, {Nair}, {Unther},
  {Deil}, {Woillez}, {Conseil}, {Kramer}, {Turner}, {Singer}, {Fox}, {Weaver},
  {Zabalza}, {Edwards}, {Azalee Bostroem}, {Burke}, {Casey}, {Crawford},
  {Dencheva}, {Ely}, {Jenness}, {Labrie}, {Lim}, {Pierfederici}, {Pontzen},
  {Ptak}, {Refsdal}, {Servillat}, \& {Streicher}}]{2013A&A...558A..33A}
{Astropy Collaboration}, {Robitaille}, T.~P., {Tollerud}, E.~J., {et~al.} 2013,
  \aap, 558, A33, \dodoi{10.1051/0004-6361/201322068}

\bibitem[{{Baron} {et~al.}(2012){Baron}, {H{\"o}flich}, {Krisciunas},
  {Dominguez}, {Khokhlov}, {Phillips}, {Suntzeff}, \&
  {Wang}}]{2012ApJ...753..105B}
{Baron}, E., {H{\"o}flich}, P., {Krisciunas}, K., {et~al.} 2012, \apj, 753,
  105, \dodoi{10.1088/0004-637X/753/2/105}

\bibitem[{{Benetti} {et~al.}(2005){Benetti}, {Cappellaro}, {Mazzali},
  {Turatto}, {Altavilla}, {Bufano}, {Elias-Rosa}, {Kotak}, {Pignata}, {Salvo},
  \& {Stanishev}}]{2005ApJ...623.1011B}
{Benetti}, S., {Cappellaro}, E., {Mazzali}, P.~A., {et~al.} 2005, \apj, 623,
  1011, \dodoi{10.1086/428608}

\bibitem[{{Blanton} \& {Roweis}(2007)}]{2007AJ....133..734B}
{Blanton}, M.~R., \& {Roweis}, S. 2007, \aj, 133, 734, \dodoi{10.1086/510127}

\bibitem[{{Branch} \& {Wheeler}(2017)}]{2017suex.book.....B}
{Branch}, D., \& {Wheeler}, J.~C. 2017, {Supernova Explosions},
  \dodoi{10.1007/978-3-662-55054-0}

\bibitem[{{Branch} {et~al.}(2006){Branch}, {Dang}, {Hall}, {Ketchum},
  {Melakayil}, {Parrent}, {Troxel}, {Casebeer}, {Jeffery}, \&
  {Baron}}]{2006PASP..118..560B}
{Branch}, D., {Dang}, L.~C., {Hall}, N., {et~al.} 2006, \pasp, 118, 560,
  \dodoi{10.1086/502778}

\bibitem[{{Brown} {et~al.}(2009){Brown}, {Holland}, {Immler}, {Milne},
  {Roming}, {Gehrels}, {Nousek}, {Panagia}, {Still}, \& {Vanden
  Berk}}]{2009AJ....137.4517B}
{Brown}, P.~J., {Holland}, S.~T., {Immler}, S., {et~al.} 2009, \aj, 137, 4517,
  \dodoi{10.1088/0004-6256/137/5/4517}

\bibitem[{{Brown} {et~al.}(2014){Brown}, {Kuin}, {Scalzo}, {Smitka}, {de
  Pasquale}, {Holland}, {Krisciunas}, {Milne}, \& {Wang}}]{2014ApJ...787...29B}
{Brown}, P.~J., {Kuin}, P., {Scalzo}, R., {et~al.} 2014, \apj, 787, 29,
  \dodoi{10.1088/0004-637X/787/1/29}

\bibitem[{{Cao} {et~al.}(2015){Cao}, {Kulkarni}, {Howell}, {Gal-Yam},
  {Kasliwal}, {Valenti}, {Johansson}, {Amanullah}, {Goobar}, {Sollerman},
  {Taddia}, {Horesh}, {Sagiv}, {Cenko}, {Nugent}, {Arcavi}, {Surace},
  {Wo{\'z}niak}, {Moody}, {Rebbapragada}, {Bue}, \&
  {Gehrels}}]{2015Natur.521..328C}
{Cao}, Y., {Kulkarni}, S.~R., {Howell}, D.~A., {et~al.} 2015, \nat, 521, 328,
  \dodoi{10.1038/nature14440}

\bibitem[{{Cao} {et~al.}(2016){Cao}, {Johansson}, {Nugent}, {Goobar}, {Nordin},
  {Kulkarni}, {Cenko}, {Fox}, {Kasliwal}, {Fremling}, {Amanullah}, {Hsiao},
  {Perley}, {Bue}, {Masci}, {Lee}, \& {Chotard}}]{2016ApJ...823..147C}
{Cao}, Y., {Johansson}, J., {Nugent}, P.~E., {et~al.} 2016, \apj, 823, 147,
  \dodoi{10.3847/0004-637X/823/2/147}

\bibitem[{{Cappellaro} {et~al.}(1997){Cappellaro}, {Mazzali}, {Benetti},
  {Danziger}, {Turatto}, {della Valle}, \& {Patat}}]{1997A&A...328..203C}
{Cappellaro}, E., {Mazzali}, P.~A., {Benetti}, S., {et~al.} 1997, \aap, 328,
  203.
\newblock \doarXiv{astro-ph/9707016}

\bibitem[{{Cardelli} {et~al.}(1989){Cardelli}, {Clayton}, \&
  {Mathis}}]{1989ApJ...345..245C}
{Cardelli}, J.~A., {Clayton}, G.~C., \& {Mathis}, J.~S. 1989, \apj, 345, 245,
  \dodoi{10.1086/167900}

\bibitem[{{Chakradhari} {et~al.}(2014){Chakradhari}, {Sahu}, {Srivastav}, \&
  {Anupama}}]{2014MNRAS.443.1663C}
{Chakradhari}, N.~K., {Sahu}, D.~K., {Srivastav}, S., \& {Anupama}, G.~C. 2014,
  \mnras, 443, 1663, \dodoi{10.1093/mnras/stu1258}

\bibitem[{{Chatzopoulos} {et~al.}(2012){Chatzopoulos}, {Wheeler}, \&
  {Vinko}}]{2012ApJ...746..121C}
{Chatzopoulos}, E., {Wheeler}, J.~C., \& {Vinko}, J. 2012, \apj, 746, 121,
  \dodoi{10.1088/0004-637X/746/2/121}

\bibitem[{{Colgate} \& {McKee}(1969)}]{1969ApJ...157..623C}
{Colgate}, S.~A., \& {McKee}, C. 1969, \apj, 157, 623, \dodoi{10.1086/150102}

\bibitem[{{Dilday} {et~al.}(2012){Dilday}, {Howell}, {Cenko}, {Silverman},
  {Nugent}, {Sullivan}, {Ben-Ami}, {Bildsten}, {Bolte}, {Endl}, {Filippenko},
  {Gnat}, {Horesh}, {Hsiao}, {Kasliwal}, {Kirkman}, {Maguire}, {Marcy},
  {Moore}, {Pan}, {Parrent}, {Podsiadlowski}, {Quimby}, {Sternberg}, {Suzuki},
  {Tytler}, {Xu}, {Bloom}, {Gal-Yam}, {Hook}, {Kulkarni}, {Law}, {Ofek},
  {Polishook}, \& {Poznanski}}]{2012Sci...337..942D}
{Dilday}, B., {Howell}, D.~A., {Cenko}, S.~B., {et~al.} 2012, Science, 337,
  942, \dodoi{10.1126/science.1219164}

\bibitem[{{Dutta} {et~al.}(2021){Dutta}, {Singh}, {Anupama}, {Sahu}, \&
  {Kumar}}]{2021MNRAS.503..896D}
{Dutta}, A., {Singh}, A., {Anupama}, G.~C., {Sahu}, D.~K., \& {Kumar}, B. 2021,
  \mnras, 503, 896, \dodoi{10.1093/mnras/stab481}

\bibitem[{{Dutta} {et~al.}(2022){Dutta}, {Sahu}, {Anupama}, {Joharle}, {Kumar},
  {Nayana}, {Singh}, {Kumar}, {Bhalerao}, \& {Barway}}]{2022ApJ...925..217D}
{Dutta}, A., {Sahu}, D.~K., {Anupama}, G.~C., {et~al.} 2022, \apj, 925, 217,
  \dodoi{10.3847/1538-4357/ac366f}

\bibitem[{{Filippenko} {et~al.}(1992){Filippenko}, {Richmond}, {Branch},
  {Gaskell}, {Herbst}, {Ford}, {Treffers}, {Matheson}, {Ho}, {Dey}, {Sargent},
  {Small}, \& {van Breugel}}]{1992AJ....104.1543F}
{Filippenko}, A.~V., {Richmond}, M.~W., {Branch}, D., {et~al.} 1992, \aj, 104,
  1543, \dodoi{10.1086/116339}

\bibitem[{{Fink} {et~al.}(2018){Fink}, {Kromer}, {Hillebrandt}, {R{\"o}pke},
  {Pakmor}, {Seitenzahl}, \& {Sim}}]{2018A&A...618A.124F}
{Fink}, M., {Kromer}, M., {Hillebrandt}, W., {et~al.} 2018, \aap, 618, A124,
  \dodoi{10.1051/0004-6361/201833475}

\bibitem[{{Foreman-Mackey} {et~al.}(2013){Foreman-Mackey}, {Conley},
  {Meierjurgen Farr}, {Hogg}, {Lang}, {Marshall}, {Price-Whelan}, {Sanders}, \&
  {Zuntz}}]{2013ascl.soft03002F}
{Foreman-Mackey}, D., {Conley}, A., {Meierjurgen Farr}, W., {et~al.} 2013,
  {emcee: The MCMC Hammer}.
\newblock \doeprint{1303.002}

\bibitem[{{Friedman} {et~al.}(2015){Friedman}, {Wood-Vasey}, {Marion},
  {Challis}, {Mandel}, {Bloom}, {Modjaz}, {Narayan}, {Hicken}, {Foley},
  {Klein}, {Starr}, {Morgan}, {Rest}, {Blake}, {Miller}, {Falco}, {Wyatt},
  {Mink}, {Skrutskie}, \& {Kirshner}}]{2015ApJS..220....9F}
{Friedman}, A.~S., {Wood-Vasey}, W.~M., {Marion}, G.~H., {et~al.} 2015, \apjs,
  220, 9, \dodoi{10.1088/0067-0049/220/1/9}

\bibitem[{{Graham} {et~al.}(2017){Graham}, {Harris}, {Fox}, {Nugent}, {Kasen},
  {Silverman}, \& {Filippenko}}]{2017ApJ...843..102G}
{Graham}, M.~L., {Harris}, C.~E., {Fox}, O.~D., {et~al.} 2017, \apj, 843, 102,
  \dodoi{10.3847/1538-4357/aa78ee}

\bibitem[{{Gurugubelli} {et~al.}(2011){Gurugubelli}, {Sahu}, {Anupama}, {Anto},
  \& {Arora}}]{2011CBET.2653....3G}
{Gurugubelli}, K., {Sahu}, D.~K., {Anupama}, G.~C., {Anto}, P., \& {Arora}, S.
  2011, Central Bureau Electronic Telegrams, 2653, 3

\bibitem[{{Hicken} {et~al.}(2007){Hicken}, {Garnavich}, {Prieto}, {Blondin},
  {DePoy}, {Kirshner}, \& {Parrent}}]{2007ApJ...669L..17H}
{Hicken}, M., {Garnavich}, P.~M., {Prieto}, J.~L., {et~al.} 2007, \apjl, 669,
  L17, \dodoi{10.1086/523301}

\bibitem[{{Hoeflich} {et~al.}(1996){Hoeflich}, {Khokhlov}, {Wheeler},
  {Phillips}, {Suntzeff}, \& {Hamuy}}]{1996ApJ...472L..81H}
{Hoeflich}, P., {Khokhlov}, A., {Wheeler}, J.~C., {et~al.} 1996, \apjl, 472,
  L81, \dodoi{10.1086/310363}

\bibitem[{{Howell} {et~al.}(2006){Howell}, {Sullivan}, {Nugent}, {Ellis},
  {Conley}, {Le Borgne}, {Carlberg}, {Guy}, {Balam}, {Basa}, {Fouchez}, {Hook},
  {Hsiao}, {Neill}, {Pain}, {Perrett}, \& {Pritchet}}]{2006Natur.443..308H}
{Howell}, D.~A., {Sullivan}, M., {Nugent}, P.~E., {et~al.} 2006, \nat, 443,
  308, \dodoi{10.1038/nature05103}

\bibitem[{{Hoyle} \& {Fowler}(1960)}]{1960ApJ...132..565H}
{Hoyle}, F., \& {Fowler}, W.~A. 1960, \apj, 132, 565, \dodoi{10.1086/146963}

\bibitem[{Hunter(2007)}]{Hunter:2007}
Hunter, J.~D. 2007, Computing in Science \& Engineering, 9, 90,
  \dodoi{10.1109/MCSE.2007.55}

\bibitem[{{Jha} {et~al.}(2019){Jha}, {Maguire}, \&
  {Sullivan}}]{2019NatAs...3..706J}
{Jha}, S.~W., {Maguire}, K., \& {Sullivan}, M. 2019, Nature Astronomy, 3, 706,
  \dodoi{10.1038/s41550-019-0858-0}

\bibitem[{{Kasen} {et~al.}(2006){Kasen}, {Thomas}, \&
  {Nugent}}]{2006ApJ...651..366K}
{Kasen}, D., {Thomas}, R.~C., \& {Nugent}, P. 2006, \apj, 651, 366,
  \dodoi{10.1086/506190}

\bibitem[{{Khatami} \& {Kasen}(2019)}]{2019ApJ...878...56K}
{Khatami}, D.~K., \& {Kasen}, D.~N. 2019, \apj, 878, 56,
  \dodoi{10.3847/1538-4357/ab1f09}

\bibitem[{{Khokhlov} {et~al.}(1993){Khokhlov}, {Mueller}, \&
  {Hoeflich}}]{1993A&A...270..223K}
{Khokhlov}, A., {Mueller}, E., \& {Hoeflich}, P. 1993, \aap, 270, 223

\bibitem[{{Krisciunas} {et~al.}(2011){Krisciunas}, {Li}, {Matheson}, {Howell},
  {Stritzinger}, {Aldering}, {Berlind}, {Calkins}, {Challis}, {Chornock},
  {Conley}, {Filippenko}, {Ganeshalingam}, {Germany}, {Gonz{\'a}lez},
  {Gooding}, {Hsiao}, {Kasen}, {Kirshner}, {Howie Marion}, {Muena}, {Nugent},
  {Phelps}, {Phillips}, {Qiu}, {Quimby}, {Rines}, {Silverman}, {Suntzeff},
  {Thomas}, \& {Wang}}]{2011AJ....142...74K}
{Krisciunas}, K., {Li}, W., {Matheson}, T., {et~al.} 2011, \aj, 142, 74,
  \dodoi{10.1088/0004-6256/142/3/74}

\bibitem[{{Kromer} {et~al.}(2013){Kromer}, {Pakmor}, {Taubenberger}, {Pignata},
  {Fink}, {R{\"o}pke}, {Seitenzahl}, {Sim}, \&
  {Hillebrandt}}]{2013ApJ...778L..18K}
{Kromer}, M., {Pakmor}, R., {Taubenberger}, S., {et~al.} 2013, \apjl, 778, L18,
  \dodoi{10.1088/2041-8205/778/1/L18}

\bibitem[{{Kushnir} \& {Katz}(2019)}]{2019RNAAS...3..162K}
{Kushnir}, D., \& {Katz}, B. 2019, Research Notes of the American Astronomical
  Society, 3, 162, \dodoi{10.3847/2515-5172/ab5064}

\bibitem[{{Kushnir} {et~al.}(2013){Kushnir}, {Katz}, {Dong}, {Livne}, \&
  {Fern{\'a}ndez}}]{2013ApJ...778L..37K}
{Kushnir}, D., {Katz}, B., {Dong}, S., {Livne}, E., \& {Fern{\'a}ndez}, R.
  2013, \apjl, 778, L37, \dodoi{10.1088/2041-8205/778/2/L37}

\bibitem[{{Langer} {et~al.}(2000){Langer}, {Deutschmann}, {Wellstein}, \&
  {H{\"o}flich}}]{2000A&A...362.1046L}
{Langer}, N., {Deutschmann}, A., {Wellstein}, S., \& {H{\"o}flich}, P. 2000,
  \aap, 362, 1046.
\newblock \doarXiv{astro-ph/0008444}

\bibitem[{{Li} {et~al.}(2003){Li}, {Filippenko}, {Chornock}, {Berger},
  {Berlind}, {Calkins}, {Challis}, {Fassnacht}, {Jha}, {Kirshner}, {Matheson},
  {Sargent}, {Simcoe}, {Smith}, \& {Squires}}]{2003PASP..115..453L}
{Li}, W., {Filippenko}, A.~V., {Chornock}, R., {et~al.} 2003, \pasp, 115, 453,
  \dodoi{10.1086/374200}

\bibitem[{{Li} {et~al.}(2011){Li}, {Bloom}, {Podsiadlowski}, {Miller}, {Cenko},
  {Jha}, {Sullivan}, {Howell}, {Nugent}, {Butler}, {Ofek}, {Kasliwal},
  {Richards}, {Stockton}, {Shih}, {Bildsten}, {Shara}, {Bibby}, {Filippenko},
  {Ganeshalingam}, {Silverman}, {Kulkarni}, {Law}, {Poznanski}, {Quimby},
  {McCully}, {Patel}, {Maguire}, \& {Shen}}]{2011Natur.480..348L}
{Li}, W., {Bloom}, J.~S., {Podsiadlowski}, P., {et~al.} 2011, \nat, 480, 348,
  \dodoi{10.1038/nature10646}

\bibitem[{{Liu} {et~al.}(2010){Liu}, {Chen}, {Wang}, \&
  {Han}}]{2010A&A...523A...3L}
{Liu}, W.~M., {Chen}, W.~C., {Wang}, B., \& {Han}, Z.~W. 2010, \aap, 523, A3,
  \dodoi{10.1051/0004-6361/201014180}

\bibitem[{{Lu} {et~al.}(2021){Lu}, {Ashall}, {Hsiao}, {Hoeflich}, {Galbany},
  {Baron}, {Phillips}, {Contreras}, {Burns}, {Suntzeff}, {Stritzinger},
  {Anais}, {Anderson}, {Brown}, {Busta}, {Castell{\'o}n}, {Davis}, {Diamond},
  {Falco}, {Gonzalez}, {Hamuy}, {Holmbo}, {Holoien}, {Krisciunas}, {Kirshner},
  {Kumar}, {Kuncarayakti}, {Marion}, {Morrell}, {Persson}, {Piro}, {Prieto},
  {Sand}, {Shahbandeh}, {Shappee}, \& {Taddia}}]{2021ApJ...920..107L}
{Lu}, J., {Ashall}, C., {Hsiao}, E.~Y., {et~al.} 2021, \apj, 920, 107,
  \dodoi{10.3847/1538-4357/ac1606}

\bibitem[{{Maeda} \& {Terada}(2016)}]{2016IJageePD..2530024M}
{Maeda}, K., \& {Terada}, Y. 2016, International Journal of Modern Physics D,
  25, 1630024, \dodoi{10.1142/S021827181630024X}

\bibitem[{{Magee} \& {Maguire}(2020)}]{2020A&A...642A.189M}
{Magee}, M.~R., \& {Maguire}, K. 2020, \aap, 642, A189,
  \dodoi{10.1051/0004-6361/202037870}

\bibitem[{{Makarov} {et~al.}(2014){Makarov}, {Prugniel}, {Terekhova},
  {Courtois}, \& {Vauglin}}]{2014A&A...570A..13M}
{Makarov}, D., {Prugniel}, P., {Terekhova}, N., {Courtois}, H., \& {Vauglin},
  I. 2014, \aap, 570, A13, \dodoi{10.1051/0004-6361/201423496}

\bibitem[{{Mazzali} {et~al.}(2007){Mazzali}, {R{\"o}pke}, {Benetti}, \&
  {Hillebrandt}}]{2007Sci...315..825M}
{Mazzali}, P.~A., {R{\"o}pke}, F.~K., {Benetti}, S., \& {Hillebrandt}, W. 2007,
  Science, 315, 825, \dodoi{10.1126/science.1136259}

\bibitem[{{Munari} \& {Renzini}(1992)}]{1992ApJ...397L..87M}
{Munari}, U., \& {Renzini}, A. 1992, \apjl, 397, L87, \dodoi{10.1086/186551}

\bibitem[{{Nugent} {et~al.}(2011){Nugent}, {Sullivan}, {Cenko}, {Thomas},
  {Kasen}, {Howell}, {Bersier}, {Bloom}, {Kulkarni}, {Kandrashoff},
  {Filippenko}, {Silverman}, {Marcy}, {Howard}, {Isaacson}, {Maguire},
  {Suzuki}, {Tarlton}, {Pan}, {Bildsten}, {Fulton}, {Parrent}, {Sand},
  {Podsiadlowski}, {Bianco}, {Dilday}, {Graham}, {Lyman}, {James}, {Kasliwal},
  {Law}, {Quimby}, {Hook}, {Walker}, {Mazzali}, {Pian}, {Ofek}, {Gal-Yam}, \&
  {Poznanski}}]{2011Natur.480..344N}
{Nugent}, P.~E., {Sullivan}, M., {Cenko}, S.~B., {et~al.} 2011, \nat, 480, 344,
  \dodoi{10.1038/nature10644}

\bibitem[{{Pakmor} {et~al.}(2010){Pakmor}, {Kromer}, {R{\"o}pke}, {Sim},
  {Ruiter}, \& {Hillebrandt}}]{2010Natur.463...61P}
{Pakmor}, R., {Kromer}, M., {R{\"o}pke}, F.~K., {et~al.} 2010, \nat, 463, 61,
  \dodoi{10.1038/nature08642}

\bibitem[{{Pakmor} {et~al.}(2012){Pakmor}, {Kromer}, {Taubenberger}, {Sim},
  {R{\"o}pke}, \& {Hillebrandt}}]{2012ApJ...747L..10P}
{Pakmor}, R., {Kromer}, M., {Taubenberger}, S., {et~al.} 2012, \apjl, 747, L10,
  \dodoi{10.1088/2041-8205/747/1/L10}

\bibitem[{pandas~development team(2020)}]{reback2020pandas}
pandas~development team, T. 2020, pandas-dev/pandas: Pandas, latest,  Zenodo,
  \dodoi{10.5281/zenodo.3509134}

\bibitem[{{Pankey}(1962)}]{1962PhDT........25P}
{Pankey}, Titus, J. 1962, PhD thesis, Howard University, Washington DC

\bibitem[{{Parrent} {et~al.}(2011){Parrent}, {Thomas}, {Fesen}, {Marion},
  {Challis}, {Garnavich}, {Milisavljevic}, {Vink{\`o}}, \&
  {Wheeler}}]{2011ApJ...732...30P}
{Parrent}, J.~T., {Thomas}, R.~C., {Fesen}, R.~A., {et~al.} 2011, \apj, 732,
  30, \dodoi{10.1088/0004-637X/732/1/30}

\bibitem[{Pedregosa {et~al.}(2011)Pedregosa, Varoquaux, Gramfort, Michel,
  Thirion, Grisel, Blondel, Prettenhofer, Weiss, Dubourg, Vanderplas, Passos,
  Cournapeau, Brucher, Perrot, \& Duchesnay}]{scikit-learn}
Pedregosa, F., Varoquaux, G., Gramfort, A., {et~al.} 2011, Journal of Machine
  Learning Research, 12, 2825

\bibitem[{{Phillips}(1993)}]{1993ApJ...413L.105P}
{Phillips}, M.~M. 1993, \apjl, 413, L105, \dodoi{10.1086/186970}

\bibitem[{{Poole} {et~al.}(2008){Poole}, {Breeveld}, {Page}, {Landsman},
  {Holland}, {Roming}, {Kuin}, {Brown}, {Gronwall}, {Hunsberger}, {Koch},
  {Mason}, {Schady}, {vanden Berk}, {Blustin}, {Boyd}, {Broos}, {Carter},
  {Chester}, {Cucchiara}, {Hancock}, {Huckle}, {Immler}, {Ivanushkina},
  {Kennedy}, {Marshall}, {Morgan}, {Pandey}, {de Pasquale}, {Smith}, \&
  {Still}}]{2008MNRAS.383..627P}
{Poole}, T.~S., {Breeveld}, A.~A., {Page}, M.~J., {et~al.} 2008, \mnras, 383,
  627, \dodoi{10.1111/j.1365-2966.2007.12563.x}

\bibitem[{{Puckett} {et~al.}(2011){Puckett}, {Newton}, {Balam}, {Bohlender},
  {Monin}, {Chisholm}, {Green}, {Gurugubelli}, {Sahu}, {Anupama}, {Anto}, \&
  {Arora}}]{2011CBET.2653....1P}
{Puckett}, T., {Newton}, J., {Balam}, D.~D., {et~al.} 2011, Central Bureau
  Electronic Telegrams, 2653, 1

\bibitem[{{Raskin} \& {Kasen}(2013)}]{2013ApJ...772....1R}
{Raskin}, C., \& {Kasen}, D. 2013, \apj, 772, 1,
  \dodoi{10.1088/0004-637X/772/1/1}

\bibitem[{{Raskin} {et~al.}(2009){Raskin}, {Timmes}, {Scannapieco}, {Diehl}, \&
  {Fryer}}]{2009MNRAS.399L.156R}
{Raskin}, C., {Timmes}, F.~X., {Scannapieco}, E., {Diehl}, S., \& {Fryer}, C.
  2009, \mnras, 399, L156, \dodoi{10.1111/j.1745-3933.2009.00743.x}

\bibitem[{Rasmussen \& Williams(2006)}]{3569}
Rasmussen, C., \& Williams, C. 2006, Gaussian Processes for Machine Learning,
  Adaptive Computation and Machine Learning (Cambridge, MA, USA: MIT Press),
  248

\bibitem[{{Rodrigo} \& {Solano}(2020)}]{2020sea..confE.182R}
{Rodrigo}, C., \& {Solano}, E. 2020, in XIV.0 Scientific Meeting (virtual) of
  the Spanish Astronomical Society, 182

\bibitem[{{Rosswog} {et~al.}(2009){Rosswog}, {Kasen}, {Guillochon}, \&
  {Ramirez-Ruiz}}]{2009ApJ...705L.128R}
{Rosswog}, S., {Kasen}, D., {Guillochon}, J., \& {Ramirez-Ruiz}, E. 2009,
  \apjl, 705, L128, \dodoi{10.1088/0004-637X/705/2/L128}

\bibitem[{{Schlafly} \& {Finkbeiner}(2011)}]{2011ApJ...737..103S}
{Schlafly}, E.~F., \& {Finkbeiner}, D.~P. 2011, \apj, 737, 103,
  \dodoi{10.1088/0004-637X/737/2/103}

\bibitem[{{Science Software Branch at STScI}(2012)}]{2012ascl.soft07011S}
{Science Software Branch at STScI}. 2012, {PyRAF: Python alternative for IRAF}.
\newblock \doeprint{1207.011}

\bibitem[{{Shen} {et~al.}(2021){Shen}, {Blondin}, {Kasen}, {Dessart},
  {Townsley}, {Boos}, \& {Hillier}}]{2021ApJ...909L..18S}
{Shen}, K.~J., {Blondin}, S., {Kasen}, D., {et~al.} 2021, \apjl, 909, L18,
  \dodoi{10.3847/2041-8213/abe69b}

\bibitem[{{Silverman} {et~al.}(2013){Silverman}, {Nugent}, {Gal-Yam},
  {Sullivan}, {Howell}, {Filippenko}, {Pan}, {Cenko}, \&
  {Hook}}]{2013ApJ...772..125S}
{Silverman}, J.~M., {Nugent}, P.~E., {Gal-Yam}, A., {et~al.} 2013, \apj, 772,
  125, \dodoi{10.1088/0004-637X/772/2/125}

\bibitem[{{Sim} {et~al.}(2010){Sim}, {R{\"o}pke}, {Hillebrandt}, {Kromer},
  {Pakmor}, {Fink}, {Ruiter}, \& {Seitenzahl}}]{2010ApJ...714L..52S}
{Sim}, S.~A., {R{\"o}pke}, F.~K., {Hillebrandt}, W., {et~al.} 2010, \apjl, 714,
  L52, \dodoi{10.1088/2041-8205/714/1/L52}

\bibitem[{{Stritzinger} {et~al.}(2006){Stritzinger}, {Mazzali}, {Sollerman}, \&
  {Benetti}}]{2006A&A...460..793S}
{Stritzinger}, M., {Mazzali}, P.~A., {Sollerman}, J., \& {Benetti}, S. 2006,
  \aap, 460, 793, \dodoi{10.1051/0004-6361:20065514}

\bibitem[{{Taubenberger}(2017)}]{2017hsn..book..317T}
{Taubenberger}, S. 2017, in Handbook of Supernovae, ed. A.~W. {Alsabti} \&
  P.~{Murdin}, 317, \dodoi{10.1007/978-3-319-21846-5\_37}

\bibitem[{{Taubenberger} {et~al.}(2013){Taubenberger}, {Kromer}, {Pakmor},
  {Pignata}, {Maeda}, {Hachinger}, {Leibundgut}, \&
  {Hillebrandt}}]{2013ApJ...775L..43T}
{Taubenberger}, S., {Kromer}, M., {Pakmor}, R., {et~al.} 2013, \apjl, 775, L43,
  \dodoi{10.1088/2041-8205/775/2/L43}

\bibitem[{{Taubenberger} {et~al.}(2011){Taubenberger}, {Benetti}, {Childress},
  {Pakmor}, {Hachinger}, {Mazzali}, {Stanishev}, {Elias-Rosa}, {Agnoletto},
  {Bufano}, {Ergon}, {Harutyunyan}, {Inserra}, {Kankare}, {Kromer},
  {Navasardyan}, {Nicolas}, {Pastorello}, {Prosperi}, {Salgado}, {Sollerman},
  {Stritzinger}, {Turatto}, {Valenti}, \& {Hillebrandt}}]{2011MNRAS.412.2735T}
{Taubenberger}, S., {Benetti}, S., {Childress}, M., {et~al.} 2011, \mnras, 412,
  2735, \dodoi{10.1111/j.1365-2966.2010.18107.x}

\bibitem[{{Tody}(1993)}]{1993ASPC...52..173T}
{Tody}, D. 1993, in Astronomical Society of the Pacific Conference Series,
  Vol.~52, Astronomical Data Analysis Software and Systems II, ed. R.~J.
  {Hanisch}, R.~J.~V. {Brissenden}, \& J.~{Barnes}, 173

\bibitem[{{Turatto} {et~al.}(2003){Turatto}, {Benetti}, \&
  {Cappellaro}}]{2003fthp.conf..200T}
{Turatto}, M., {Benetti}, S., \& {Cappellaro}, E. 2003, in From Twilight to
  Highlight: The Physics of Supernovae, ed. W.~{Hillebrandt} \&
  B.~{Leibundgut}, 200, \dodoi{10.1007/10828549\_26}

\bibitem[{{van den Heuvel} {et~al.}(1992){van den Heuvel}, {Bhattacharya},
  {Nomoto}, \& {Rappaport}}]{1992A&A...262...97V}
{van den Heuvel}, E.~P.~J., {Bhattacharya}, D., {Nomoto}, K., \& {Rappaport},
  S.~A. 1992, \aap, 262, 97

\bibitem[{Van Der~Walt {et~al.}(2011)Van Der~Walt, Colbert, \&
  Varoquaux}]{van2011numpy}
Van Der~Walt, S., Colbert, S.~C., \& Varoquaux, G. 2011, Computing in Science
  \& Engineering, 13, 22

\bibitem[{{Virtanen} {et~al.}(2020){Virtanen}, {Gommers}, {Oliphant},
  {Haberland}, {Reddy}, {Cournapeau}, {Burovski}, {Peterson}, {Weckesser},
  {Bright}, {van der Walt}, {Brett}, {Wilson}, {Jarrod Millman}, {Mayorov},
  {Nelson}, {Jones}, {Kern}, {Larson}, {Carey}, {Polat}, {Feng}, {Moore}, {Vand
  erPlas}, {Laxalde}, {Perktold}, {Cimrman}, {Henriksen}, {Quintero}, {Harris},
  {Archibald}, {Ribeiro}, {Pedregosa}, {van Mulbregt}, \&
  {Contributors}}]{2020SciPy-NMeth}
{Virtanen}, P., {Gommers}, R., {Oliphant}, T.~E., {et~al.} 2020, Nature
  Methods, 17, 261, \dodoi{https://doi.org/10.1038/s41592-019-0686-2}

\bibitem[{{Wang} {et~al.}(2009{\natexlab{a}}){Wang}, {Li}, {Filippenko},
  {Foley}, {Kirshner}, {Modjaz}, {Bloom}, {Brown}, {Carter}, {Friedman},
  {Gal-Yam}, {Ganeshalingam}, {Hicken}, {Krisciunas}, {Milne}, {Silverman},
  {Suntzeff}, {Wood-Vasey}, {Cenko}, {Challis}, {Fox}, {Kirkman}, {Li}, {Li},
  {Malkan}, {Moore}, {Reitzel}, {Rich}, {Serduke}, {Shang}, {Steele}, {Swift},
  {Tao}, {Wong}, \& {Zhang}}]{2009ApJ...697..380W}
{Wang}, X., {Li}, W., {Filippenko}, A.~V., {et~al.} 2009{\natexlab{a}}, \apj,
  697, 380, \dodoi{10.1088/0004-637X/697/1/380}

\bibitem[{{Wang} {et~al.}(2009{\natexlab{b}}){Wang}, {Filippenko},
  {Ganeshalingam}, {Li}, {Silverman}, {Wang}, {Chornock}, {Foley}, {Gates},
  {Macomber}, {Serduke}, {Steele}, \& {Wong}}]{2009ApJ...699L.139W}
{Wang}, X., {Filippenko}, A.~V., {Ganeshalingam}, M., {et~al.}
  2009{\natexlab{b}}, \apjl, 699, L139, \dodoi{10.1088/0004-637X/699/2/L139}

\bibitem[{{Yaron} \& {Gal-Yam}(2012)}]{2012PASP..124..668Y}
{Yaron}, O., \& {Gal-Yam}, A. 2012, \pasp, 124, 668, \dodoi{10.1086/666656}

\bibitem[{{Yoon} \& {Langer}(2005)}]{2005A&A...435..967Y}
{Yoon}, S.~C., \& {Langer}, N. 2005, \aap, 435, 967,
  \dodoi{10.1051/0004-6361:20042542}

\end{thebibliography}

\end{document}